\documentclass[a4paper]{jpconf}

\usepackage{hyperref}
\usepackage{graphicx}

\begin{document}

 
\newcommand{\cern}{CERN}
\newcommand{\cvmo}{CernVM Online}
\newcommand{\cvmcg}{CernVM Cloud Gateway}
\newcommand{\cernvm}{CernVM}
\newcommand{\cvmfs}{CernVM-FS}


\title{\cernvm\ Online and Cloud Gateway: a uniform interface for \cernvm\ contextualization and deployment}

\author{G Lestaris$^1$, I Charalampidis$^2$, D Berzano, J Blomer, P Buncic, G Ganis and R Meusel}
\address{CERN, CH-1211 Geneva 23, Switzerland}
\ead{$^1$george.lestaris@cern.ch}
\ead{$^2$ioannis.charalampidis@cern.ch}

\begin{abstract}
In a virtualized environment, contextualization is the process of configuring a VM instance for the needs of various deployment use cases. Contextualization in \cernvm\ can be done by passing a handwritten context to the user data field of cloud APIs, when running \cernvm\ on the cloud, or by using \cernvm\ web interface when running the VM locally. \cvmo\ is a publicly accessible web interface that unifies these two procedures. A user is able to define, store and share \cernvm\ contexts using \cvmo\ and then apply them either in a cloud by using \cvmcg\, or on a local VM with the single-step pairing mechanism. \cvmcg\ is a distributed system that provides a single interface to use multiple and different clouds (by location or type, private or public). Cloud gateway has been so far integrated with OpenNebula, CloudStack and EC2 tools interfaces. A user, with access to a number of clouds, can run \cernvm\ cloud agents that will communicate with these clouds using their interfaces, and then use one single interface to deploy and scale \cernvm\ clusters. \cernvm\ clusters are defined in \cvmo\ and consist of a set of \cernvm\ instances that are contextualized and can communicate with each other. 
\end{abstract}


\section{Introduction}\label{sec:introduction}

At \cernvm\ \cite{1742-6596-219-4-042003}, \emph{contextualization} is the process of applying a predefined set of parameters to a \emph{polymorphic} virtual machine instance, converting it to a specialized processing unit. This process removes the need of distributing large OS templates for every different processing unit configuration.

By the term \emph{polymorphic} we mean a virtual machine that provides all basic components required, but they are not configured nor specialized. This means that an instance of a \emph{polymorphic} VM can be configured to become any kind of processing unit.

This concept is implemented in CernVM 2.x series with the image flavors. There are four different kind of \cernvm\ images that target four different kinds of flavours: \cernvm\ \emph{Head Node}, \cernvm\ \emph{Batch Node}, \cernvm\ \emph{Basic} and \cernvm\ \emph{Desktop}. The first two are designed to be used in cluster environments. The \emph{Head Node} contains an HTCondor \cite{condor05} master and monitoring utilities while \emph{Batch Node} contains an HTCondor client. \cernvm\ \emph{Basic} contains most of the widely-used components; \cernvm\ \emph{Desktop} is the same as \emph{Basic}, with the addition of a graphical interface.


However with the introduction of \cernvm\ 3.x series \cite{microCernVM} there will be a single image covering all the use cases. This increases the flexibility inherited from \cernvm\ 2.x since, with \cernvm\ 3.x, contextualization can transform a newly booted virtual machine to either \emph{batch head}/\emph{node} or \emph{desktop}.

There are several ways a \cernvm\ image can be contextualized.

\subsection{Automatic contextualization}

\subsubsection{Amiconfig}

The contextualization process is applied using a patched version of \emph{amiconfig} \cite{cernvm-amiconfig-web} from \emph{rPath}. This tool was designed to download user-data provided via an EC2 infrastructure and apply them to the system. In \cernvm\ we have modified this tool in order to accept configuration from other sources and in order to handle more configuration cases.


In detail, \emph{amiconfig} is a tool written in Python and has a pluggable architecture. It is composed of a downloading class and a set of plugins that apply the configuration fetched. 

\subsubsection{Passing user-data}

User-data can be provided to a \cernvm\ instance using various techniques. Currently \cernvm\ supports EC2 compatible method, CloudStack and HEPIX standards. These methods are used by the vast majority of cloud providers. Also with \cernvm\ 3.x, cloud-init \cite{cloudinit-web} package is included. When AMIconfig fails to find the context, with one of the previously mentioned methods, cloud-init will be used and its various data sources will be applied.


\subsection{Web appliance: manual contextualization}

For \cernvm\ instances that are deployed locally, earlier described contextualization is still possible with the \emph{HEPIX compliant CD-ROM}. However, \cernvm\ provides also a user friendlier approach aiming to help inexperienced users. This is the \emph{web appliance} that comes with \cernvm\ image and starts once the machine is booted. User can access the web appliance by using a URL that is printed in VM's console.

In the remaining of this paper we describe \cvmo\ in section \tref{sec:cvmo}, a web portal to define, store and share contexts, as well as easier ways to easily apply them in local virtual machines and virtual machines with console access. Moreover we introduce a model for formulating virtual clusters in section \ref{sec:vc-model}, and \cvmcg\ in section \ref{sec:cvmcg}, a distributed system for virtual cluster deployment.


\section{\cvmo}\label{sec:cvmo}

\begin{figure}[h]
	\begin{center}
		\includegraphics[width=0.8\textwidth]{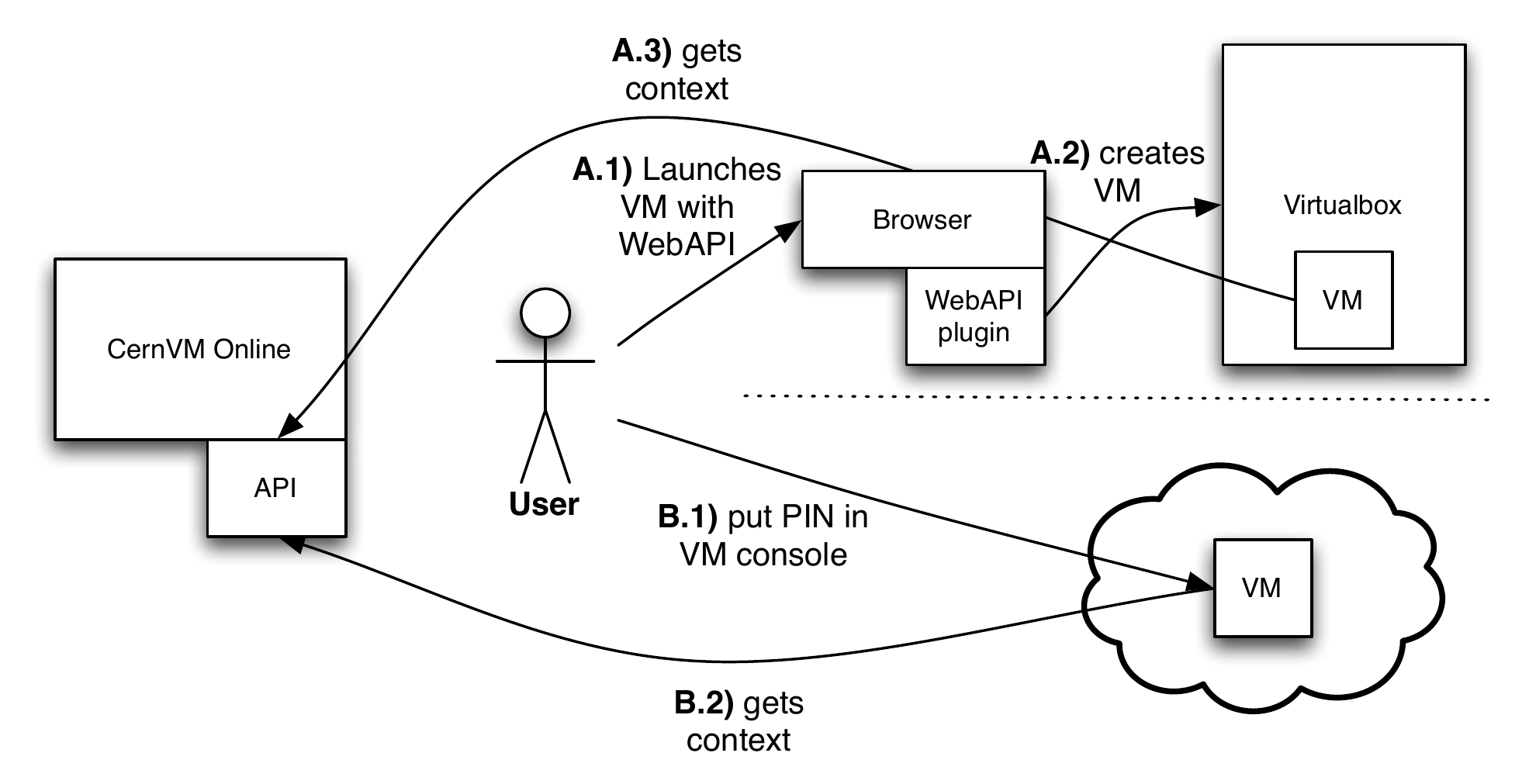}
	\end{center}
	\caption{\cvmo\ use cases}
	\label{fig:cvmo-workflows}
\end{figure}

\cvmo\ is a web portal that can be considered as an \emph{online database} of contextualization information. It provides a simplified interface that allows end-users to create, archive and share contextualization information or -- called for short -- \emph{contexts}. \cvmo\ is also the endpoint that every \cernvm\ instance will contact in order to fetch the appropriate contextualization information. It provides two methods to initiate this procedure: pairing and integration with WebAPI which can be seen in figure \ref{fig:cvmo-workflows} and are described in the following sections.

In addition to per-machine contextualization information, \cvmo\ provides management capabilities for \emph{cluster definitions}.  A cluster definition is a virtual representation of a cluster. Virtual cluster definition is described in section \ref{sec:vc-model}.

This service is currently hosted at \url{http://cernvm-online.cern.ch}. You can log-in with your \cern\ credentials, or you can create an external account. Note however that this is a public service and does not target only \cern\ users.

\subsection{Context definition}

A context is the file that contains all the information required to configure a \cernvm\ instance. Contexts in \cvmo\ are created using an interface that follows the same principle of dividing the context into sections that provide configuration parameters for different contextualization plugins. Hence the interface itself allows for selecting which plugins will be enabled and defining their settings.

\subsubsection{Immutable contexts}

Contexts stored in \cvmo\ are immutable. User is not allowed to modify them. This is done to keep consistency of stored context and context already applied on a VM instance. Nevertheless, if the user needs to modify a context, \cvmo\ generates a new context (with a new id) that inherits all the settings from the ``parent'' context and provides user with an interface to make required modifications. This procedure is called \emph{cloning}.

\subsubsection{Security}

Contexts sometimes contain sensitive information such us passwords and keys. \cvmo\ provides the \emph{encrypted context} feature that enables encryption of the context contents  with a user defined pass phrase. The contents are stored encrypted in the database and any access to the context (for cloning and pairing) requires the context password.

\subsubsection{Marketplace}

\cvmo\ is also a place for sharing contexts for common use cases within the user community. A user that defines a context for a common purpose can \emph{publish} it to the \cvmo\ \emph{marketplace}. The contexts are then organised into categories, indexed by tags and users can clone and pair them.

\subsection{Pairing}

\cvmo\ contexts can be extracted in AMI config format using the web interface or the API. This solves to problem of manually writing the contexts. Additionally the \emph{pairing} mechanism enables contextualization of running virtual machines by just typing a PIN number.

Pairing requires access to VM's console. There the user can specify the pairing PIN obtained from \cvmo. The VM will contact \cvmo\ to get the context and it will start the contextualization process to apply it. The VM information such us name, \cernvm\ version and IP address will be stored in \cvmo\ and the user is able to get a list of paired virtual machines.

\subsection{Integration with WebAPI}

\emph{\cernvm\ WebAPI} browser plugin has been developed to deploy local \cernvm\ instances from the browser window using the VirtualBox hypervisor. It supports Internet explorer, Firefox and Chrome and it has been integrated experimentally with \cvmo. This integration allows user to deploy a \cernvm\ 3 instance by selecting the context and the instance hardware configuration (CPU, memory, disk). The WebAPI plugin will install VirtualBox in the system (if it is not already there), download the tiny $\mu$\cernvm\ image and start the machine. Once the machine boots it will be contextualized with the selected context.


\section{Modeling virtual clusters}\label{sec:vc-model}

A virtual cluster is a computing cluster that consists solely by \emph{virtual machines}. Clusters contain machines with \emph{different roles}. Each role is defined by the services a machine runs. However machine's role also defines the configuration of the machine itself. There are machines in a cluster which are more important and that all the other machines have to connect to. These machines require special contextualization, different hardware configuration, and in strict networks different firewall settings.

The model developed to define virtual clusters separates these different machine types based on their role. This distinction is known by the name \emph{services}. Each service in our virtual cluster can be implemented by a single VM or by many. There are two types of services based on their scalability behaviour. The \emph{fixed services} and the \emph{scalable services}.

\begin{figure}[h]
	\begin{center}
		\includegraphics[width=0.3\textwidth]{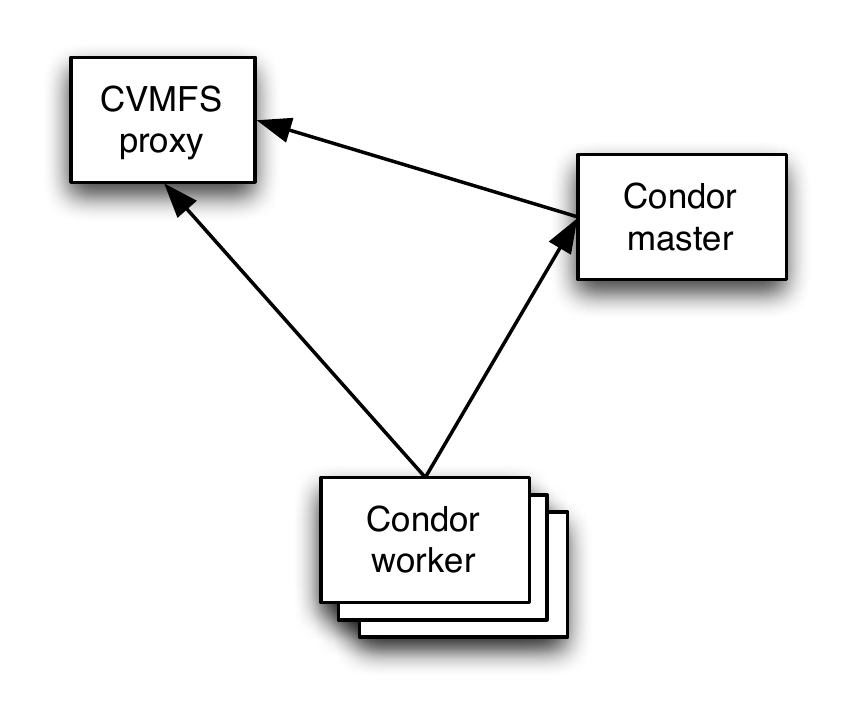}
	\end{center}
	\caption{A typical virtual cluster}
	\label{fig:cvmcg-vc-model}
\end{figure}

\paragraph{Fixed services} They are services that are deployed once for each virtual cluster instance. A fixed service has a fixed number of instances implementing it. In batch systems the fixed service would be the head node. Other examples of fixed services are proxies and VO boxes.

\paragraph{Scalable service} These are services that are been implemented by a varying number of VMs. As the name suggests, these services can be scaled up or down. Typical example is the worker nodes of a batch system.

\subsection{Service dependancies}

Usually the scalable service instances communicate with the fixed service instances. This implies that the fixed service instances must be available when a scalable service instance is created. To formulate the order of VM deployment we use service dependancies. Each service can depend to a number of services that must be deployed for it to start. The dependancies also used to destroy VMs. Once a fixed service instance is not used (required) by any other VM, it can be destroyed.

In the current implementation of \cvmo\ interface for defining virtual clusters, dependancies are set automatically by the order of services. Also all the scalable services depend on all the fixed services. This interface will be extended in the future to allow to define more flexible dependancies between services.

\subsection{Service definition}

A service requires, and is identified by, a \emph{name}. Also it describes the software and hardware configuration of its instances. \emph{Software configuration} is selected by defining the \emph{\cvmo\ context} and \emph{\cernvm\ image/version} to use for a service. \emph{Hardware configuration} is defined by the concept of offerings. There are three types of offerings: \emph{compute}, \emph{disk} and \emph{network} offerings.

\section{\cvmcg}\label{sec:cvmcg}

\cvmcg\ is a distributed system used to spawn virtual clusters in multiple and different clouds. Its front end is a web server, the \emph{cloud gateway server}, that provides the interface to the users as well as a REST API. The cloud gateway server generates requests that are put into the shared database that \emph{gateway agents} poll. The the gateway agents communicate with the remote \emph{cloud agents} (one for each connected cloud) through XMPP.

\begin{figure}[h]
	\begin{center}
		\includegraphics[width=0.6\textwidth]{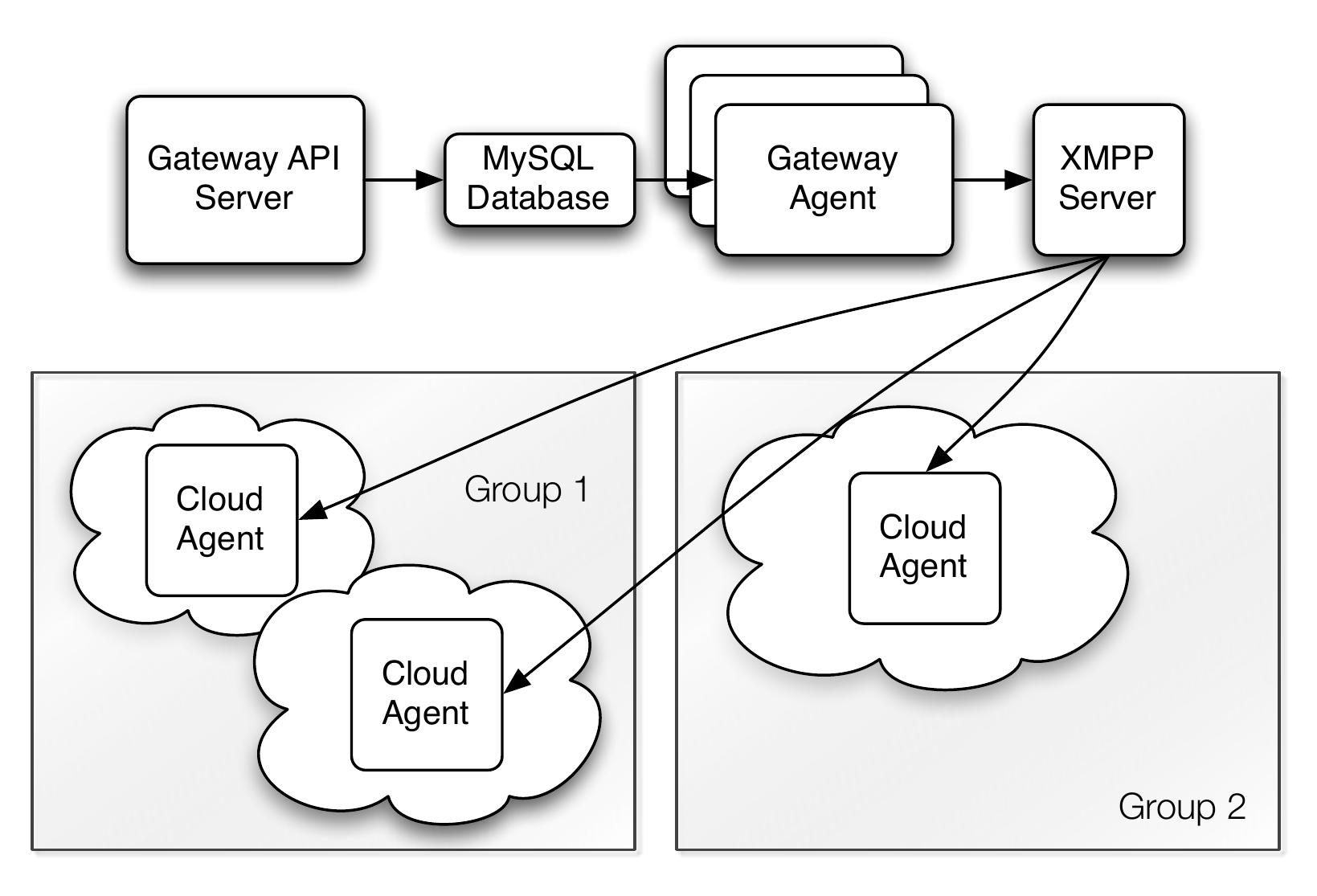}
	\end{center}
	\caption{The architecture of the \cvmcg}
	\label{fig:cvmcg-arch}
\end{figure}

More information on the \cvmcg\, API documentation, implementation details and administrator reference can be found in the technical report \cite{cvmc-tr} of the project.

\subsection{Cloud gateway server}

The cloud gateway server provides the user interface and the API of the system. Access requires authentication which is based on \cvmcg\ accounts for the user interface and API  credentials for the API. Note that a user can create and use more than one set of API credentials and revoke them independently. The provided interface allows for:

\begin{itemize}
	\item Creation of new clusters, according to a cluster definition stored in \cvmo
	\item Scale up cluster services
	\item Pause, resume or destroy individual instances and
	\item Pause, resume or destroy the entire cluster
\end{itemize}

The gateway server is using a MySQL database to keep the state of the system, the deployed clusters, their virtual machines, the connected clouds, the users and groups, etc. Though all the above mentioned actions required communication with the actual cloud APIs. This communication is asynchronous and it is handled by the gateway agent.

\subsection{Gateway agent}

All the asynchronous requests are stored in the database. One or more (for scaling) gateway agents are polling this database for new requests. Once a new request is made, a gateway agent will process it accordingly.

The gateway agent will need to communicate with connected clouds to process the requests. In the case of a cluster creation or scaling up, the gateway agent will speak to more than one clouds to select those that will be used. For virtual machine specific operation (pause, resume, destroy) the gateway agent will need to speak directly with the cloud that runs this virtual machine.

The communication with the clouds is done through cloud agents which are remote processes. Gateway agents and cloud agents are both implemented with the iAgent framework and they communicate through XMPP PubSub channels and directed messages.

\subsection{Cloud agent}

A cloud agent is the process that will listen to requests from gateway agents and that will eventually make the API calls to the connected cloud. In contrast to gateway agents, cloud agents can speak with a single cloud.

Since cloud agent is a remote process it can be deployed and hosted by the cloud administrator himself. Thus, in order to contribute cloud resources to \cvmcg\ no cloud API credentials need to be stored centrally. The administrator of the cloud agent is able to define soft quotas for limiting the used resources as well as ACL for defining which users and groups of \cvmcg\ will be able to use a certain cloud.

Finally the administrator has to define the mapping between the globally defined images, and offerings to cloud specific images and VM flavors. The cloud agent is able to communicate with different cloud APIs as this communication is handled by a module called \emph{cloud driver} which provides a uniform interface to different cloud APIs.

\subsection{iAgent framework}

iAgent framework \cite{1742-6596-396-3-032022} is a Perl framework that is used to develop distributed, agent-based systems that communicate through XMPP. The concept is similar to the actors model. Each agent is a process that has its own configuration. It requires XMPP credentials and defines programmatically accessible endpoints that expose its functionality. 

iAgent is based on POE Perl library. That allows us to define hooks that are used for message passing between the modules of the agent. Again, each module exports its functionality and iAgent kernel calls for message passing are invoking its callbacks.

\subsection{Overflowing clouds}

\begin{figure}[h]
	\begin{center}
		\includegraphics[width=0.5\textwidth]{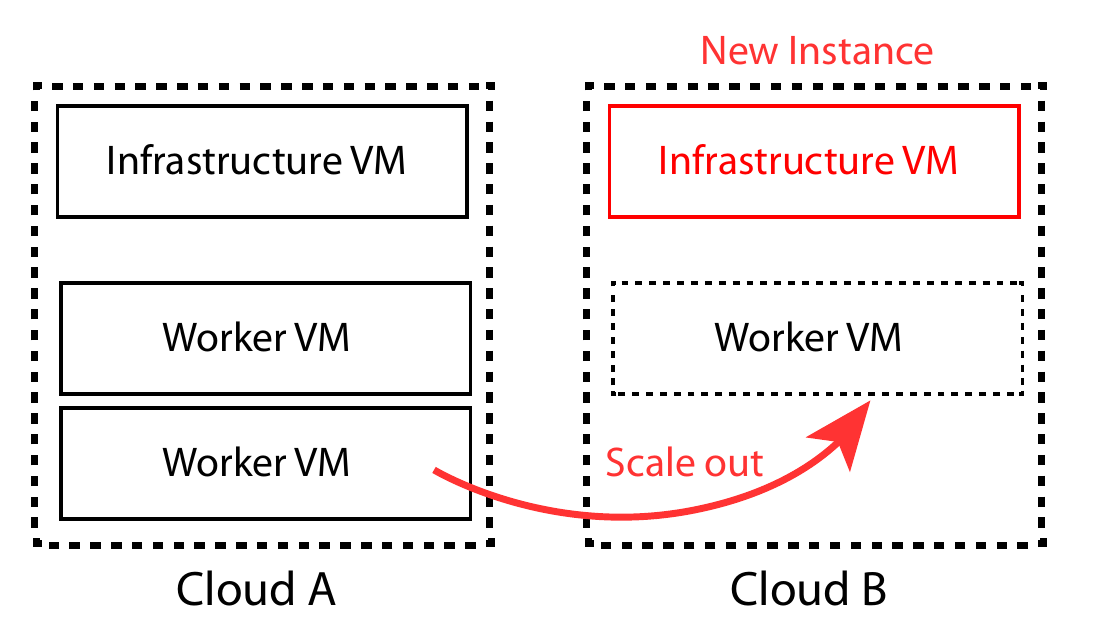}
		\caption{Scaling a cluster beyond the capabilities of the current host cloud}
	\end{center}
\end{figure}

\cvmcg\ is able to use multiple clouds. Also the definition of a virtual cluster is based on services and dependencies. These two facts are related and combined into \cvmcg\ feature of scaling up a cluster even if it does not fit to the current used cloud, by expanding it to different cloud(s). The communication between cluster services though has to be ensured and since inter cloud networking might be slow or even not possible, \cvmcg\ provides all the required services to the scaled service, in the new cloud.

Service replication ensures that the service that needs scaling will be able to locally connect to other required services while it will not be limited to the cloud that was selected in the cluster deployment. Also this enables the usage of multiple cloud at the same time and for the same cluster.


\section{Conclusion}\label{sec:conclusion}

We presented the developments in \cernvm\ contextualization processes and the introduction of \cvmo\ web portal for managing and sharing contexts. Additionally we described the methods provided to apply a context to local virtual machines, deployed on user's PC, or to \cernvm\ instances that are already running and where user has console access.

Then we proposed a model for defining virtual clusters by separating types of machines into services, having fixed and scalable services as well as setting service dependancies. We showed the \cvmcg\ which uses this model to deploy and contextualise virtual clusters based on \cernvm\ instances over multiple and different clouds.

Finally we briefly mentioned the \cvmcg\ feature of overflowing clouds by making use of our flexible model for defining virtual cluster services dependancies and by replicating required services for a cluster instance to new clouds. The later is certainly something worth exploring further to enable the simultaneous use of different, heterogenous and geographically distributed cloud resources.


\section*{References}
\bibliography{proceedings}

\end{document}